\documentclass[12pt]{iopart}
\usepackage{graphicx}
\usepackage{epsfig}
\usepackage{graphics}
\usepackage{bm}
\usepackage{psfrag}
\usepackage[english]{babel}
\usepackage[T1]{fontenc}
\usepackage{color}
\usepackage[dvipsnames]{xcolor}
\usepackage{braket}

\usepackage{umoline}

\usepackage[colorlinks,bookmarks=false,citecolor=blue,linkcolor=red,urlcolor=black]{hyperref}
\usepackage[T1]{fontenc}
\usepackage{epsfig}

\def\urlprefix{}
\def\url#1{}
\def\be{\begin{equation}}
\def\ee{\end{equation}}
\def\bea{\begin{eqnarray}}
\def\eea{\end{eqnarray}}
\def\bi{\begin{itemize}}
\def\ei{\end{itemize}}
\def\bin{\begin{enumerate}}
\def\ein{\end{enumerate}}

\def\bg{\begin{equation}\begin{gathered}}
\def\eg{\end{gathered}\end{equation}}

\begin{document}
\title{Many-body localization of bosons in optical lattices}
\author{Piotr Sierant$^{1}$ and Jakub Zakrzewski$^{1,2}$}
\ead{jakub.zakrzewski@uj.edu.pl}
\address{$^1$ Instytut Fizyki imienia Mariana
Smoluchowskiego, Uniwersytet Jagiello\'nski, ulica \L{}ojasiewicza
11, PL-30-348 Krak\'ow, Poland.} 
\address{$^2$ Mark Kac Complex
Systems Research Center, Uniwersytet Jagiello\'nski, Krak\'ow,
Poland. }

\date{\today}

\begin{abstract}
 Many--body localization for a system of bosons trapped in a one dimensional lattice is discussed. 
 Two models that may be realized for cold atoms in
optical lattices are considered.   The  model with a random 
on-site potential is compared with previously 
introduced random interactions model. While the origin and character of the disorder in both systems 
is different they show interesting similar properties. In particular, many--body localization appears 
for a sufficiently large disorder as verified by a time evolution of   
initial density wave  states as well as using statistical properties of energy levels for small system
sizes. Starting with different initial states, we observe that the localization properties are 
energy--dependent which reveals an inverted
many--body localization edge in both systems (that finding is also verified by statistical analysis 
of energy spectrum). Moreover, we consider computationally challenging regime
of  transition between many body localized and extended phases where we observe a characteristic 
algebraic decay of density correlations which may be attributed to 
subdiffusion (and Griffiths--like regions) in the studied systems. Ergodicity breaking in the 
disordered Bose--Hubbard models is compared with the
slowing--down of the time evolution of the clean system at large interactions.
\end{abstract}
\maketitle

\section{Introduction}
The effects of interactions on disordered localized physical systems remained to a large extent a mystery  
for over fifty years after the pioneering work of Anderson \cite{Anderson58} who introduced the concept of
single--particle localization.
The study of interactions in the metallic regime practically began with the work of Altshuler and Aronov
\cite{Altshuler79},
subsequently problems related to the presence
of the disorder and interactions were addressed by several works (a highly incomplete list may include 
\cite{Altshuler80,Fukuyama80,Fleishman1980,Shepelyansky94}) also in cold atomic settings \cite{Damski2003}.
It was in the seminal paper \cite{Basko06} that many body localization (MBL) was identified as a genuine new
phenomenon occurring for
 sufficiently strong disorders. This stimulated numerous studies of various aspects of the 
 MBL in the last ten years (for reviews see  
\cite{Huse14,Rahul15,Abanin17,Alet17}). 
Presently, it is a common understanding that
MBL is the most robust way of 
ergodicity breaking in the quantum world. 

Most of the theoretical studies of MBL {were} performed  for interacting spin models in a lattice, 
(e.g. Heisenberg, XXZ) as amply reviewed
in e.g. \cite{Huse14,Rahul15,Abanin17,Alet17}. Those spin models were frequently mapped on spinless fermions. 
Experimentally, both fermionic
\cite{Schreiber15,Kondov15} as well as bosonic species \cite{White09,Choi16} were investigated 
where the latter seem to be particularly challenging.
While for 1/2-spins (spinless fermions) an on--site Hilbert space is two dimensional (and for spinful fermions 
four dimensional), for bosons we have to effectively deal with  
much larger dimensions of local Hilbert space (constrained, strictly speaking, by the total number of particles, $N$) 
unless we want to consider the artificial case of hard-core bosons 
\cite{Tang15}. This makes bosonic systems unique.

In this work we consider MBL in the Bose--Hubbard model due to two 
distinct mechanisms -- either resulting from  random interactions (we extend here our previous 
studies \cite{Sierant17, Sierant17b}) or from random on--site potential. 
Bosonic systems have the advantage of being easily controlled and prepared in an experiment.
Moreover, 
the local Hilbert space is unconstrained in the bosonic case as mentioned above. That provides additional 
freedom in the choice of initial states and on one hand provides the experimentalist
with supplementary ways of studying ergodicity breaking and on the other hand leads to more complex dynamics
that makes numerical simulations of bosonic systems
a challenging task.

Consider the standard Bose--Hubbard model in one dimension
 \begin{equation}
  \displaystyle{
 H = -J \sum_{\langle i,j \rangle} \hat{a}^{\dag}_i\hat{a}_j +
 \frac{ U }{2} \sum_i  \hat{n}_i (\hat{n}_i - 1) +
 \mu \sum_i \hat{n}_i.
  } 
  \nonumber
  \end{equation}
There are two straightforward and experimentally feasible ways of introducing a genuine random
disorder to the system. The Bose--Hubbard model with random on--site potential can be simulated by an optical 
speckle field (assuming that the correlation length of the speckle is smaller than the lattice spacing):
 \begin{equation}
  \displaystyle{
 H = -J \sum_{\langle i,j \rangle} \hat{a}^{\dag}_i\hat{a}_j +
 \frac{ U }{2} \sum_i  \hat{n}_i (\hat{n}_i - 1) +
  \sum_i  \mu_i \hat{n}_i
   \label{eq: BH_random_potential}
  } 
  \end{equation}
 with  $\mu_i \in [-W/2,\, W/2]$. While this seems quite straightforward, a careful analysis
 \cite{Zhou10} shows that a realistic tight-binding model
 for a random speckle potential imposed on top of an optical lattice leads to a more 
 sophisticated model with all the parameters being random
 and drawn from speckle potential induced distributions. Nevertheless, later we will restrict 
 ourselves to the model  (\ref{eq: BH_random_potential}) as a
 plausible simplification of the experimental situation. Especially
 as the state of 
 the art imaging under a quantum microscope \cite{Bakr2009} enables a
 creation of an arbitrary optical potential by appropriate holographic masks.
  
  The second option is realized when the optical lattice  is placed close to 
  an atom chip which provides a spatially random magnetic 
field, that,  in the vicinity of  Feschbach resonance, leads to random interactions \cite{Gimperlein05}
 \begin{equation}
  \displaystyle{
 H = -J \sum_{\langle i,j \rangle} \hat{a}^{\dag}_i\hat{a}_j +
 \sum_i  \frac{U_i}{2} \hat{n}_i (\hat{n}_i - 1)+
  \sum_i  \mu \hat{n}_i
  } 
  \label{eq: BH_random_interactions}
  \end{equation}
  $U_i \in [0,\, U]$.
  The latter system (\ref{eq: BH_random_interactions}) was shown in the proceeding work \cite{Sierant17} to be 
  many--body localized at sufficiently large interaction strength amplitudes $U$.  
  More careful analysis of the level statistics of the random interactions model
  and a preliminary observation of an inverted mobility edge were presented in  \cite{Sierant17b}.
  
Let us mention also that  quasi-periodic potentials are used in fermionic experiments \cite{Schreiber15,Lueschen17} 
and  analyzed
  theoretically - for most recent works see \cite{Ancilotto18,Zakrzewski18}. 
  We restrict ourselves to an analysis of
 models with random disorder, uniformly distributed in the respective intervals.

  The two systems (\ref{eq: BH_random_potential}) and (\ref{eq: BH_random_interactions}) are clearly physically different.
  In the absence of interactions, eigenstates of (\ref{eq: BH_random_potential}) are localized whereas 
  the single--particle spectrum of (\ref{eq: BH_random_interactions}) consists of Bloch--waves and is thus fully
  delocalized. However, as we shall show below, both models behave, to a large 
  extent, quite similarly in the presence of the strong disorder. On one side this
  suggests, that MBL is a robust phenomenon  -- importantly its signatures are 
  also similar for fermionic and pure spin systems. On the other side,
  the remaining differences point towards system specific phenomena that may be studied.

  Let us set the energy and time scales by putting $J=1$ (also $\hbar=1$). The average properties of 
  the system with random 
  interactions are dependent on just 
  a single scale, $U$. The features of the system with a random on--site potential are a result of an interplay
  between the disorder characterized by its strength $W$ and interactions $U$. Therefore, one cannot identify 
  disorder strengths at which the properties of the two systems would be exactly the same.
 
 The  paper is structured as follows. First, in Section~\ref{secimbal} we analyze ergodicity breaking in the 
 disordered Bose--Hubbard 
 systems by examining the relaxation properties of specifically prepared density--wave like initial states during 
 the time evolution. Above certain disorder strengths, both systems fail to relax to a uniform density profile.
 Quantifying the degree of ergodicity breaking, we observe that it crucially depends on the energy of the initial
 state which 
  leads us to the study of the level statistics of the systems in an energy--resolved way and 
  allows us to postulate the existence of the localization edge in the system in Section~\ref{seclocal}. 
 We provide {qualitative} arguments supporting the existence of the localization edge in the
 perturbative
 language in Section~\ref{pertu}.
 We consider in a detailed way (Section~\ref{statu}) the level spacings of the random 
 on--site potential model showing that they are consistent
 with {our} imbalance studies. Finally, in Section~\ref{fateu},
 many--body localization in the disordered system is contrasted with the 
 slow down of thermalization in the clean system \cite{Carleo12}.
 
\section{The imbalance decay}
\label{secimbal}

The Eigenstate Thermalization Hypothesis (ETH) \cite{Deutsch91, Srednicki94,Cohen16}, states that  excited 
eigenstates of an ergodic system
have thermal expectation values of physical observables. In effect, a time average of a local 
observable equilibrates
to the microcanonical average 
and remains near that value for most of the time.
In order to investigate ergodicity breaking in the considered systems, we adapt a strategy analogous 
to the experiment \cite{Schreiber15} and study
the time evolution of highly excited out--of--equilibrium states. For fermions the standard interaction
mechanism is related to collisions of opposite spin fermions at a given site.
Similarly, for the interaction between bosons  to take place
one needs at least two bosons at a single site.
Therefore, as the initial state we typically consider the
 density wave-like Fock state $|DW_{21}\rangle = |2121...\rangle$ (working at 3/2 filling).
 Quantitative results are obtained via the measurement of the imbalance
 \begin{equation}
  \displaystyle{
 I(t) = \frac{D(t)}{D(0)} \quad \mathrm{with} \quad D(t)=N_e(t)-N_o(t),
  } 
  \label{eq: imbalance}
  \end{equation}
where $N_{e,o}$ are populations of even and odd sites of the 1D lattice respectively. 
A non--zero stationary value of imbalance at large times shows that the system breaks ergodicity, which, in the context of
an interacting many--body quantum system implies that it is many--body localized.

\begin{figure}[ht]
 \begin{center}
\includegraphics[width=0.8\columnwidth]{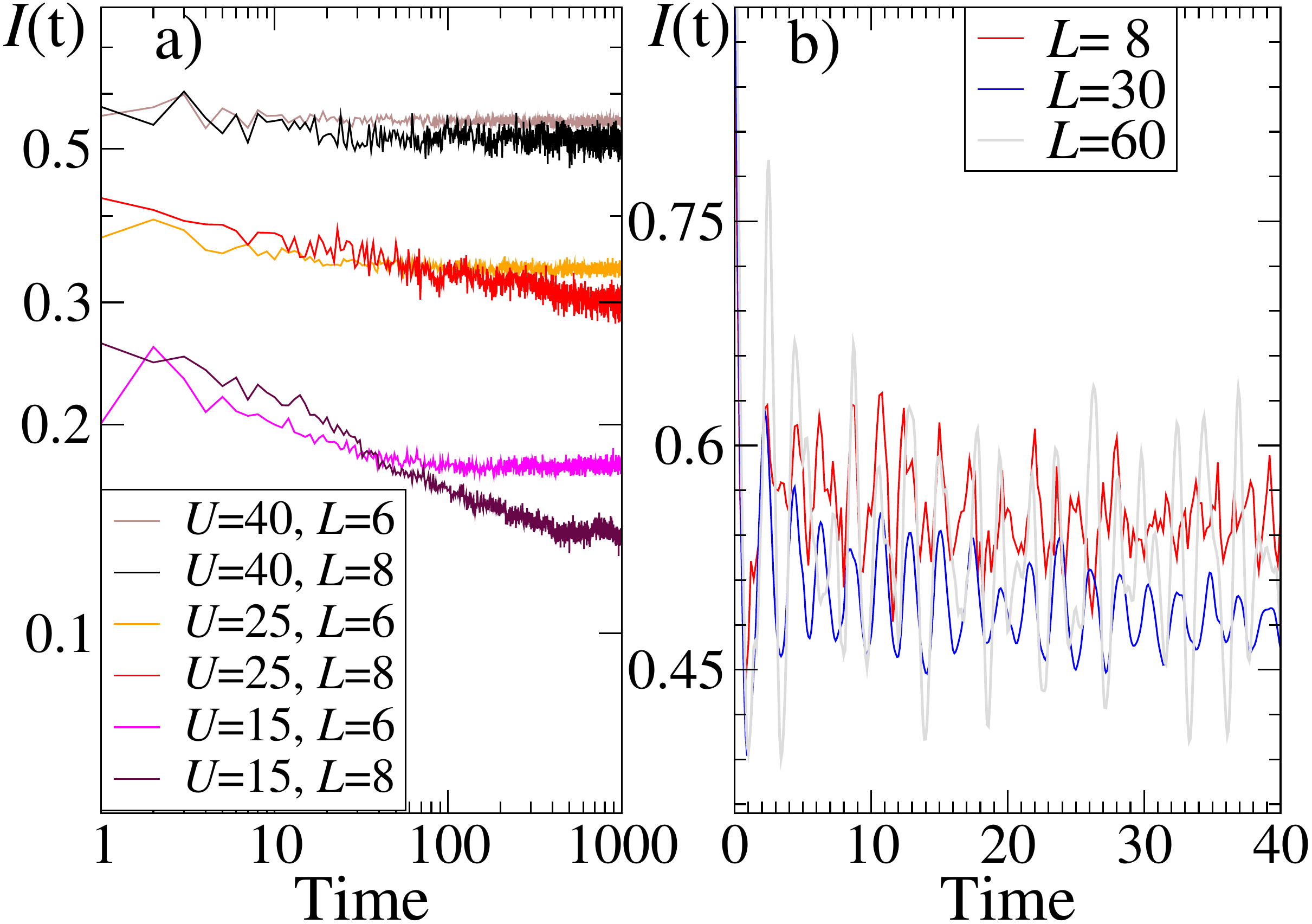}
\end{center}
\caption{ \label{fig: imb_finite_size}
Decay of $I(t)$ for $|DW_{21}\rangle$ state as a function of time. Results for the random interactions system
(\ref{eq: BH_random_interactions}) with $N=9$ and $N=12$
bosons on $L=6$ and $L=8$ sites respectively on the left. The log-log scale facilitates observation of
the emergence 
and breakdown (at times $t$ larger than the Heisenberg time $T_H$) of the algebraic decay of the imbalance $I(t)$.
Right -- comparison between $L=8,30,60$ lattice sizes for the random potential system (\ref{eq: BH_random_potential})
at $U=1$. 
}
\label{Heisenberg}
\end{figure}

\begin{figure}[ht]
 \begin{center}
\includegraphics[width=0.8\columnwidth]{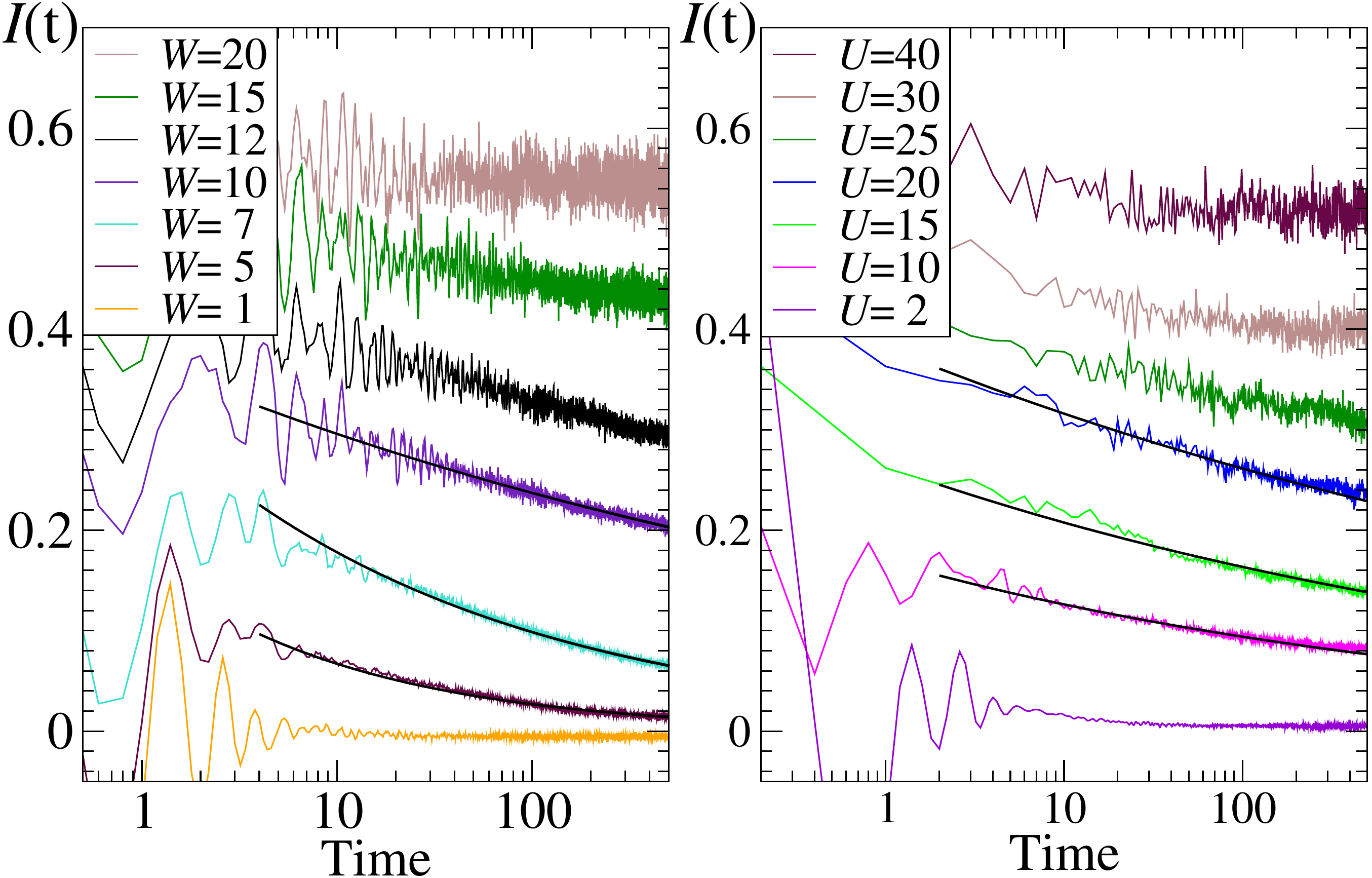}
\end{center}
\caption{ \label{fig: imbTL8N12}Decay of $I(t)$ for $|DW_{21}\rangle$ state as a function of time 
for $N=12$ bosons on $L=8$ lattice sites
-- random on--site potential with $U=1$ on the left and random 
interactions on the right. The horizontal axis is in logarithmic scale which 
facilitates observation of slow decay of $I(t)$ in 
the intermediate regime between
many--body localized and the ergodic phases. Data corresponding to the critical region are fitted by power--law decay.
}
\label{fig: imb1}
\end{figure}

Two complementary numerical tools may be used to study the time evolution of the 
system starting from a high energy nonstationary state (as e.g. $|DW_{21}\rangle$). 
For small system sizes one may use the exact diagonalization approach and 
 calculate the evolution operator after finding eigenvalues
and eigenvectors of the problem. Such an approach has been used for (less demanding computationally) 
spin systems \cite{Luitz15,Luitz16} and allows one to reach long time dynamics. 
 A bit larger systems may be quite efficiently emulated  
using time propagation in  Krylov subspaces \cite{Park86} 
- the method often referred to as the Lanczos approach, however, 
the calculations become more computationally demanding with the 
increasing evolution time. 

For finite size systems the Heisenberg time $T_H$ (being essentially the inverse of the mean level spacings) 
sets a time scale at which the time evolution freezes.
This is illustrated in Fig.~\ref{Heisenberg}a where the  decay of the imbalance $I(t)$ for the 
system of $N=9$ bosons on $L=6$ 
lattice sites with interaction strength amplitude $U=15$ effectively stops at $T_H \approx 70$ (the dimension
of the Hilbert space is equal to $2002$). However, already 
for $L=8$ the Hilbert space dimension is $50388$ and the Heisenberg time is $T_H\approx 500$ which 
allows for a clear observation of the decay of the 
imbalance $I(t)$. Fig.~\ref{Heisenberg}b demonstrates the effect of the system size and role of the 
number of disorder realizations $n$. { Data for $L=8$ and $L=30$ show that the imbalance $I(t)$, 
after an initial transient, 
 oscillates around a certain stationary value which decreases slightly 
for larger system sizes (also visible in 
Fig.~\ref{Heisenberg}a for $U=40$)}.
Moreover, the amplitude of the oscillations decreases like $1/\sqrt{n}$
with number of disorder realizations being $n=50$ for $L=8,30$ and $n=10$ for $L=60$.

The obtained time evolution of the imbalance
averaged over $50$ disorder realizations for different disorder strengths is
presented in Fig.~\ref{fig: imbTL8N12} for L=12 bosons on N=8 sites with open boundary conditions. 
At small disorder strengths both systems obey ETH and the density pattern of the $|DW_{21}\rangle$ state relaxes to uniform 
density -- this is the case for $W=1$ (left panel) and $U=2$ (right panel). For large disorder, on the other hand, 
after a rapid initial decay,
the imbalance saturates at a disorder strength dependent non-zero value showing quite significant 
fluctuations in time (as well as between 
different realizations of the disorder - not shown). The non-zero value of the long-time imbalance
indicates that the system remembers its initial state which  
is the commonly used indicator of  the MBL phase. In the broad transition 
regime between the two phases the decay of $I(t)$ is well fitted
by an algebraic decay  $I(t) \propto t^{-\frac{1}{z}}$ with the exponent $\frac{1}{z}$  decreasing  with the increasing 
disorder strength (the decay slows down). This is a similar behavior to that observed in fermionic and spin 
systems \cite{Agarwal15, Lev15, Torres-Herrera15, Luitz16} as well as experimentally for spinful fermions \cite{Lueschen17}.
In this region, which we call a quantum critical region \cite{Vosk2015,Potter2015,Khemani17}, transport is 
claimed to be subdiffusive and dominated by Griffiths--type dynamics \cite{Griffiths69,Vojta10}. 
According to the Griffiths phase model of the MBL transition \cite{Vosk2015,Potter2015,Agarwal16} $z$ is the dynamical 
exponent associated with the transport which reaches $\frac{1}{z} =0.5$ in the diffusive limit \cite{Lueschen17}. 
As the border of MBL phase is approached, the exponent $\frac{1}{z}$ vanishes.
Let us stress that both bosonic systems, despite a richer local Hilbert space than for spin/fermion models, share very
similar subdiffusive characteristics with, e.g. $\frac{1}{z}\approx 0.4$ for $W=5$ while 
$\frac{1}{z}\approx 0.09$ for $W=10$. Let us also note that the size of this quantum 
critical region is system size dependent as discussed in detail e.g. in \cite{Khemani17}.

\begin{figure}[ht]
\begin{center}
\includegraphics[width=0.8\columnwidth]{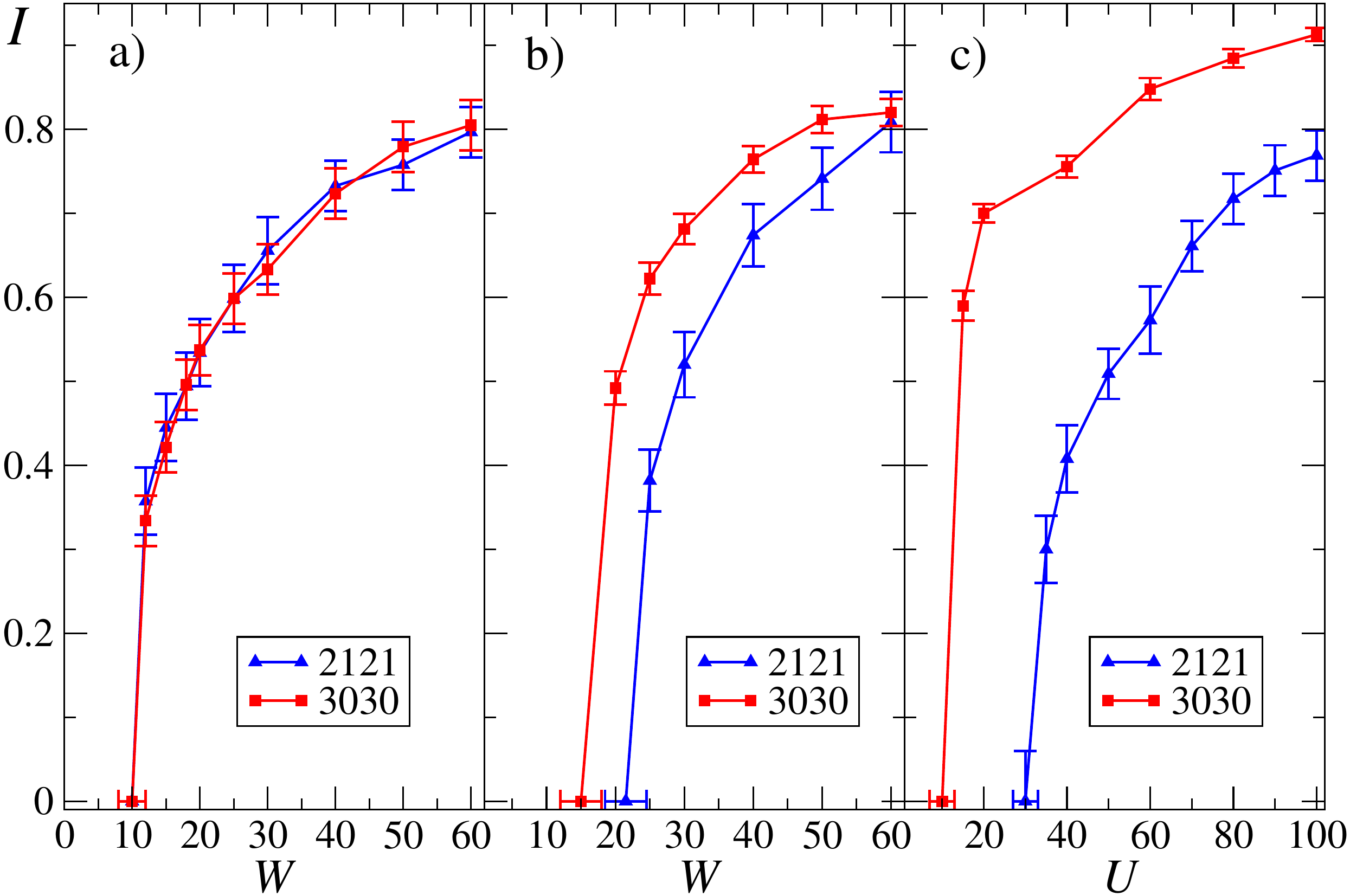}
\end{center}
\caption{ \label{fig: imbL60N90}  Stationary value $I$ of imbalance as a function of the disorder strength for 
 for two  initial Fock states: $|DW_{21}\rangle$ and $|DW_{30}\rangle$ for 90 bosons on 60 lattice sites. 
 Panel a) corresponds to a random chemical potential
in a model of  weakly interacting bosons $U=1$. Within statistical errors   both states have similar
threshold value of non-zero imbalance around $W=10$ and show essentially the same behavior.
Panel b) corresponds to the same model with stronger interactions $U=5$. Now the state localize 
at different value of the disorder amplitude. For random interactions an even stronger dependence on the 
initial state is observed -- panel c). 
}
\end{figure}

The analysis with Lanczos time propagation due to the exponentially increasing Hilbert space size is necessarily 
restricted to moderate system sizes. On the other hand, tDMRG \cite{Vidal03,Vidal04,Zakrzewski09,Schollwoeck11} 
 allows one to efficiently study  systems with a moderate growth of the entanglement in time - a situation expected 
 for localized and close to localized systems.
In particular, it is well known that the 
entanglement entropy in the MBL phase for initially uncorrelated 
parts grows logarithmically in time \cite{Serbyn13a, Huse14},
which allows us to study very large systems
\cite{Sierant17}. Here, we consider  the
time--evolution close to the MBL phase for 90 bosons distributed between $L=60$ sites 
(a typical size for cold-atoms experiments)
to infer properties of the localized system as well as to get a glimpse of dynamics near the MBL 
transition -- Fig. \ref{fig: imbL60N90}. In all tDMRG calculations 
we use open boundary conditions as realized in quasi-one-dimensional experimental situations.

The stationary value of imbalance can be thought of -- in analogy to an order parameter in conventional phase transitions
-- as a quantity
which determines the degree of ergodicity breaking in the system. It is straightforward to obtain the stationary 
imbalance deep in the many--body localized phase -- performing time evolution with tDMRG we observe that $I(t)$ 
after initial transient (up to $t\approx 10$) 
saturates at a disorder strength dependent level exhibiting
residual fluctuations in time. Therefore, we average $I(t)$ over a time interval in the vicinity of $t=30$. The 
stationary value of the
obtained imbalance is further averaged over (typically ten) disorder realizations.
 
As the disorder strength decreases, the subdiffusive regime with the algebraic decay of the imbalance  is approached.
This is accompanied with a much faster 
build up of the entanglement which, in turn, reduces the obtainable final propagation time but, on the other hand,
confirms the vicinity of the transition.
The representative stationary values of the imbalance are presented in Fig.~\ref{fig: imbL60N90} for $|DW_{21}\rangle$
as well as a different
density like state $|DW_{30}\rangle= |3030...\rangle$. Observe a striking difference between the left and
the middle panels of Fig.~\ref{fig: imbL60N90} that correspond to different interaction strengths $U=1$ and $U=5$, 
respectively. The fate of both initial states for $U=1$ is very similar, they begin to show non--zero stationary imbalance
around the amplitude of random chemical potential $W=10$. $I(t)$ dependence on $W$ for both states is practically 
identical (compare error bars). The situation drastically changes for $U=5$ where $|DW_{30}\rangle$ state localizes for
much lower values of $W$. This may be attributed (and will be further confirmed in the next Section) to the significant 
difference in initial energies of both states.

Imbalances obtained for the model with random interactions are shown in  Fig.~\ref{fig: imbL60N90}c.
As for the $U=5$ case discussed above, the $|DW_{30}\rangle$ initial state leads to a larger value of the 
stationary imbalance than the $|DW_{21}\rangle$ for a given disorder strength.
Also, the disorder strength needed to obtain the non--zero stationary value of 
imbalance is smaller for the 
$|DW_{30}\rangle$ state than for the $|DW_{21}\rangle$ state. The $|DW_{30}\rangle$ state has higher 
 energy  than $|DW_{21}\rangle$ state
as the interaction term grows quadratically with the number of bosons occupying a lattice site. 
Thus, also for this model  the degree of ergodicity breaking 
 depends on the energy of the initial state. 
 The energy dependence of the MBL phenomenon in both systems leads us to the next section where we
 examine the properties of the full energy spectra by exact diagonalization.

\section{Localization edge}
\label{seclocal}

The theory of  metal-insulator transition \cite{Mott} implies that the mobility (localization) 
edge separates in energy localized and extended states, at least in the thermodynamic limit.
For interacting systems in the presence of disorder, numerical evidence for the presence of 
many--body mobility edges were presented for
the random field Heisenberg spin chain \cite{Luitz15} and for the 
fermionic Hubbard system \cite{Naldesi16,Lin17}. 

To address the properties of the system as a function of energy in a systematic way we follow  \cite{Luitz15} and 
analyze the statistics of energy eigenvalues using a convenient measure -  the gap ratio \cite{Oganesyan07}.
Let $\delta_n$ be a difference between adjacent energies in the ordered spectrum,
 $\delta_n = E_{n+1} - E_n.$
     The  (dimensionless) gap ratio is defined as:
    \begin{equation}
        r_n = \min\{\delta_n,\delta_{n-1}\} / \max\{\delta_n,\delta_{n-1}\}.
    \end{equation}
It has been shown   \cite{Oganesyan07} that the mean of the  $r_n$ distribution, $\overline{r}$, 
describes the character of the eigenstates well. For delocalized disordered states one intuitively
expects a situation resembling random matrices. For time reversal invariant systems, 
the Gaussian Orthogonal Ensemble (GOE) of random matrices is appropriate.
 In this case, the mean gap ratio, $\overline{r}$, can be calculated approximately \cite{Atas13} yielding  
 $\overline r_{GOE} = 0.53$. In the opposite case, deep in the MBL phase, it is conjectured 
 that the system  is integrable and can be characterized by a complete set of local integrals of 
 motion (LIOMs)
 \cite{Serbyn13b,Huse14,Imbrie17}. As such, the spectrum should share properties with the Poisson ensemble, with 
 the mean ratio equal to $\overline r_{Poisson} = 2\log 2 -1 \approx 0.39$ \cite{Atas13}. 
 In the transition regime between these two phases one expects that the mean ratio has 
 intermediate values smoothly interpolating between the limiting cases. Such a situation 
 has been indeed observed in a number of studies \cite{Oganesyan07,Pal10,Luitz15,Luitz16,Sierant17}.

We shall characterize the studied systems with the help of $\overline r$ in an energy resolved way. 
To  that end, for each value 
of the disorder strength (being $W$ or $U$  depending on the studied system), 
we collect more than $10^6$ 
eigenvalues for about 200 disorder realizations. We drop the lowest 1\% of energy levels, let us denote 
the lower bound
of the set of eigenvalues as $E_{bot}$. Similarly, we disregard the highest 1\% of eigenvalues and define as
$E_{top}$ the upper bound of the considered set.
We rescale the members of the set as $\epsilon=(E-E_{bot})/(E_{top}-E_{bot})$ mapping the energies onto 
$[0,1]$. Finally, we group $\epsilon$ values in 20 equal intervals
and find $\overline r$ in each of them separately. 

\begin{figure*}[ht]
\includegraphics[width=0.33\columnwidth]{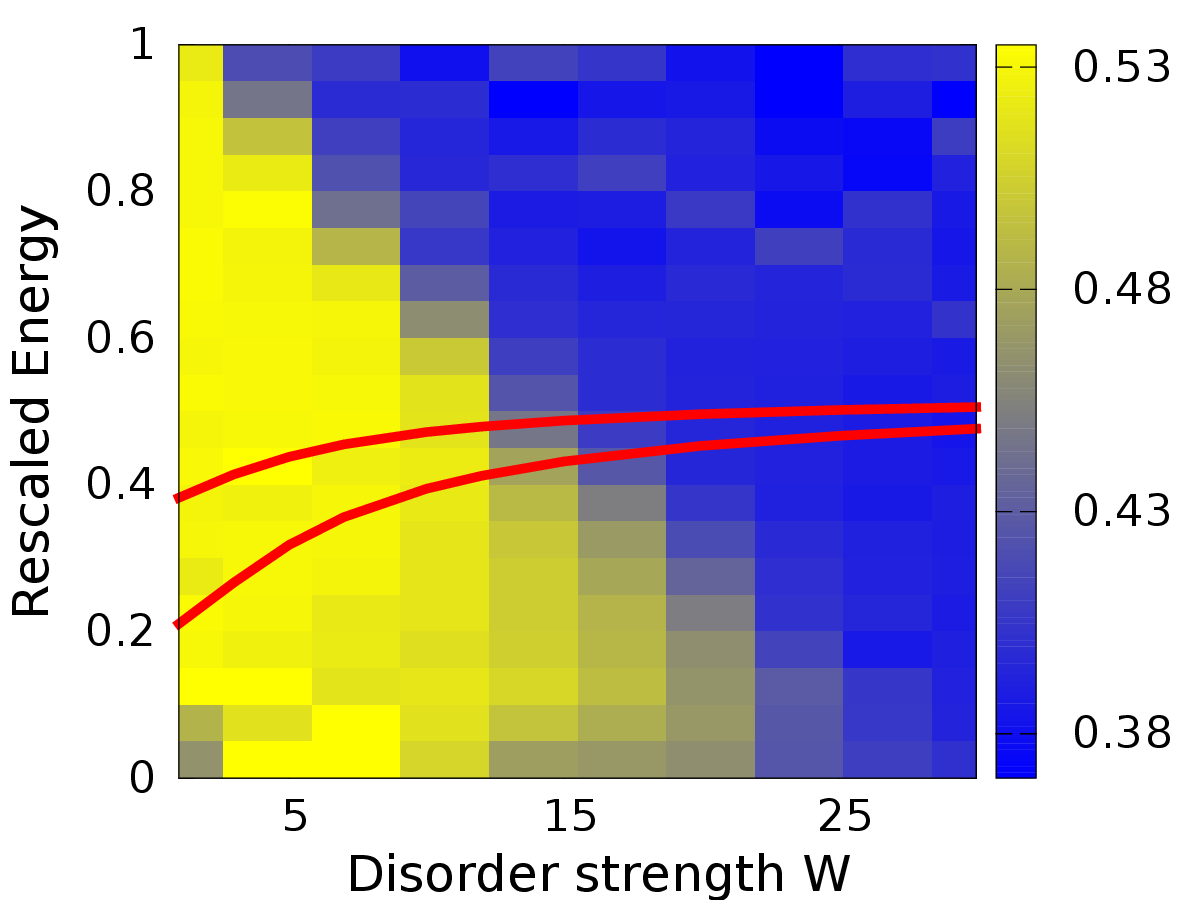}\includegraphics[width=0.33\columnwidth]{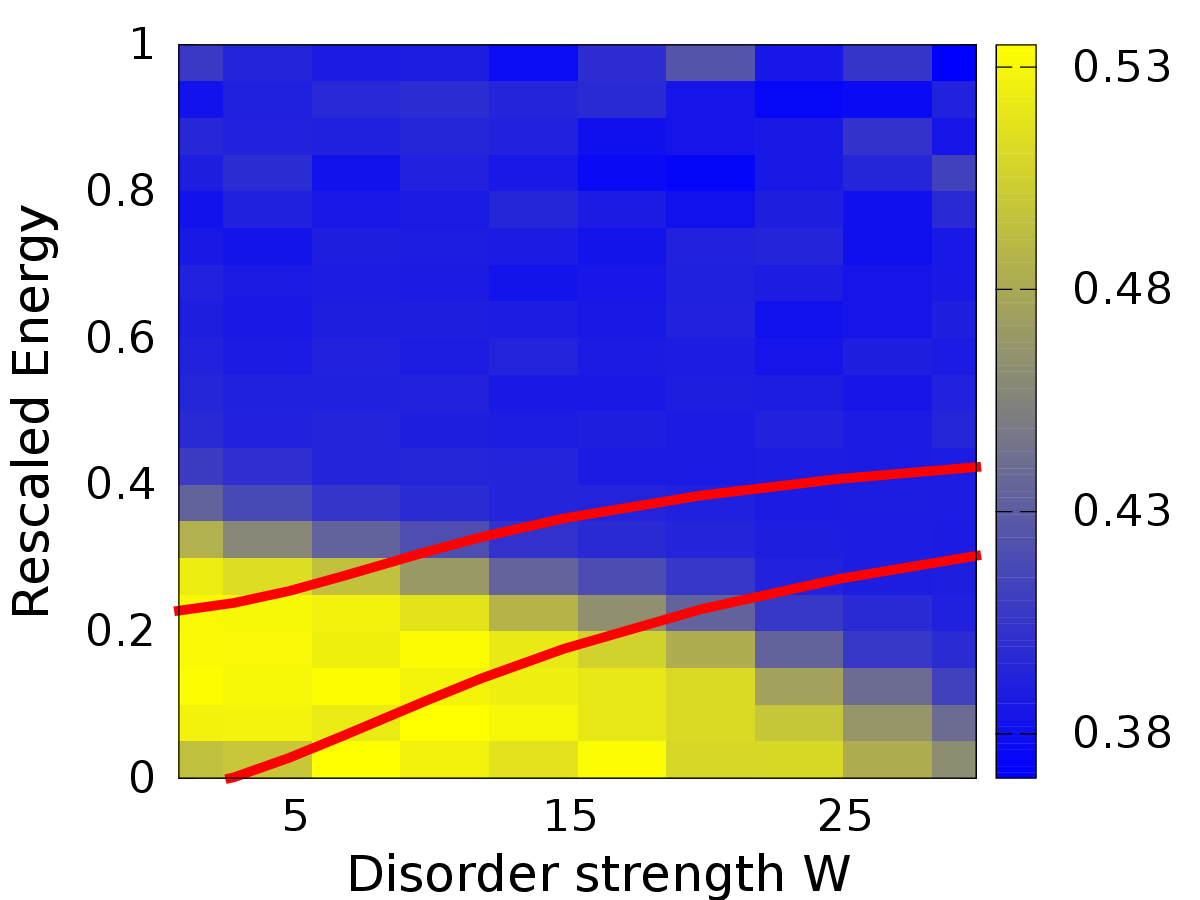}\includegraphics[width=0.33\columnwidth]{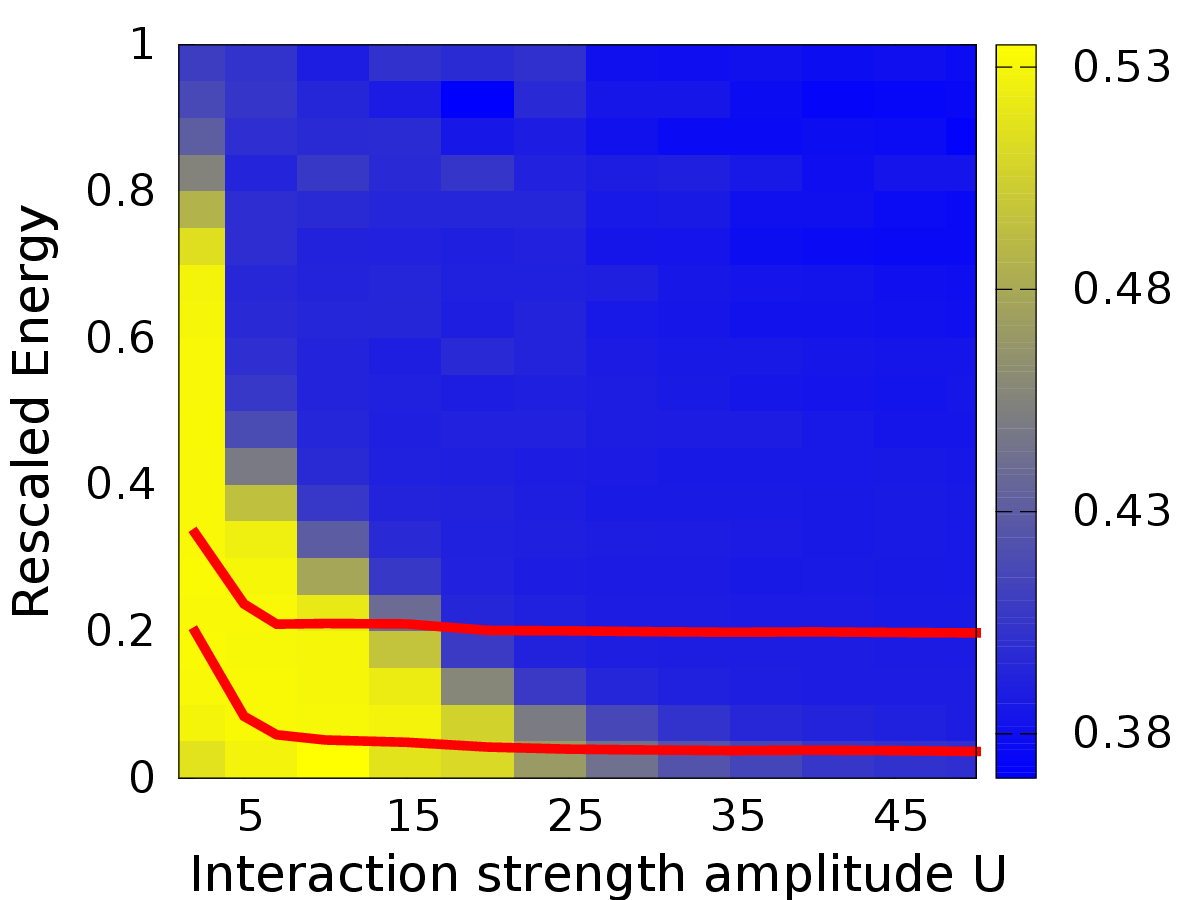}
\caption{ \label{mob1}The mean ratio of consecutive spacing in the plane of disorder strength
($W$ or $U$ depending on the model) and rescaled energy  $\epsilon = \frac{E-E_{bot}}{E_{top} - E_{bot} }$
with  $N=12$ bosons on $L=8$ sites. Left panel - random chemical potential  for $U=1$;
middle panel - the same system for $U=5$; 
right panel - the  random interactions case.
Yellow color corresponds to $\overline{r} \approx 0.53$ and to the ergodic regime whereas the blue 
color denotes $\overline{r} \approx 0.39$ characteristic
for localized states. Red curves indicate energies of the $|DW_{21}\rangle$ and $|DW_{30}\rangle$
states which cross the boundary between ergodic and localized
regions of the spectrum with increasing disorder strength and exhibit ergodicity breaking.
}
\end{figure*}

The results are presented in Fig.~\ref{mob1}. Clearly, all the systems 
 reveal a transition between the ergodic phase (yellow color)
with GOE-like $\overline{r}$ values and the localized (MBL) phase with statistics close to  
the Poisson limit. Typically,
for a broad range of disorder strengths one may notice that, for a given disorder strength,  
high lying energy 
states are localized
while the lowest remain extended. Thus, there exists an
interval of energies 
(for a given disorder strength) where a transition from localized to extended states takes place. Such a behaviour is 
typically associated with the mobility edge for single-particle systems. Thus, we may loosely say that
the apparent mobility edge is indeed observed for the studied systems. In both cases, it has a peculiar feature 
that states which are above a certain energy threshold are localized 
whereas the states below are extended, so it may be called an ``inverted'' mobility edge.  This behavior 
has already been predicted for a two-- and few--site bosonic 
systems in \cite{Singh17} with random on--site potentials.

While the transition between localized and extended states as a function of energy is clearly  observed, it is by no means 
obvious that a sharp mobility edge exists in the thermodynamic limit. 
The properties of the transition regime (defined by the energy 
interval in which $\overline{r}$ takes intermediate values between Poisson and GOE limits)
may be attributed to a mixture of localized and extended states for any finite size of the system or could stem from fractal properties 
of eigenstates. 
Another option is that the situation may be similar to the transition
between chaotic and regular (integrable) motion in simple chaotic systems (see e.g. \cite{Haake}) where ``regular'' 
states localized in the stable islands may coexist with chaotic eigenstates. In the deep semiclassical limit the residual 
tunnelings between regular islands and the chaotic sea decay exponentially (as $\hbar\rightarrow 0$) and regular and 
chaotic states may co-exist (leading e.g. to the so called Berry-Robnik statistics of levels  \cite{Prozen94}). 
The character of states in the transition regime is not known and whether the transition ``sharpens'' in the 
thermodynamic limit leading to a true mobility edge is an open question 
as the ergodic-MBL transition is claimed to be dynamical in nature \cite{Khemani17}. 
Also, the disorder is known to smear the transition due to the presence of Griffiths regions \cite{Vojta10}.

We observe that  regions of extended (yellow) and localized (blue) behavior have different shapes depending on 
the model as well as on the parameters -- see Fig~\ref{mob1}.
The left ($U=1$) and middle ($U=5$) panels correspond to the random on-site potential model.
For stronger interactions 
extended states are limited to the lower part of the energy spectrum. High energy states, necessarily having 
significant occupations on selected sites, are localized even for very small disorder. 
This difference in shape for different strengths of interactions  nicely correlates with  different temporal
behaviors of the $|DW_{21}\rangle$ and $|DW_{30}\rangle$ imbalances. 
The energies of these states are represented by red 
 lines in all three panels of Fig~\ref{mob1}. 
 They are quite similar during the transition from the extended to the localized 
  regime in the left panel ($U=1$), thus they should have a similar stationary imbalance. 
 Indeed, this is the case as shown in the left panel of 
 Fig.~\ref{fig: imbL60N90}. For stronger
 interactions, the two red lines enter the localized (blue) region for different disorder strengths 
 -- as faithfully reproduced by the
 the disorder dependence of stationary imbalance (middle panel in  Fig.~\ref{fig: imbL60N90}).
 Similar correlation is observed for the random interactions model (right panels of Fig~\ref{mob1}
 and Fig.~\ref{fig: imbL60N90}).
 
 In the next section we provide perturbative arguments backing our conclusion on observation 
 of the apparent inverted mobility edge.

\begin{figure}[ht]
\begin{center}
\includegraphics[width=0.8\columnwidth]{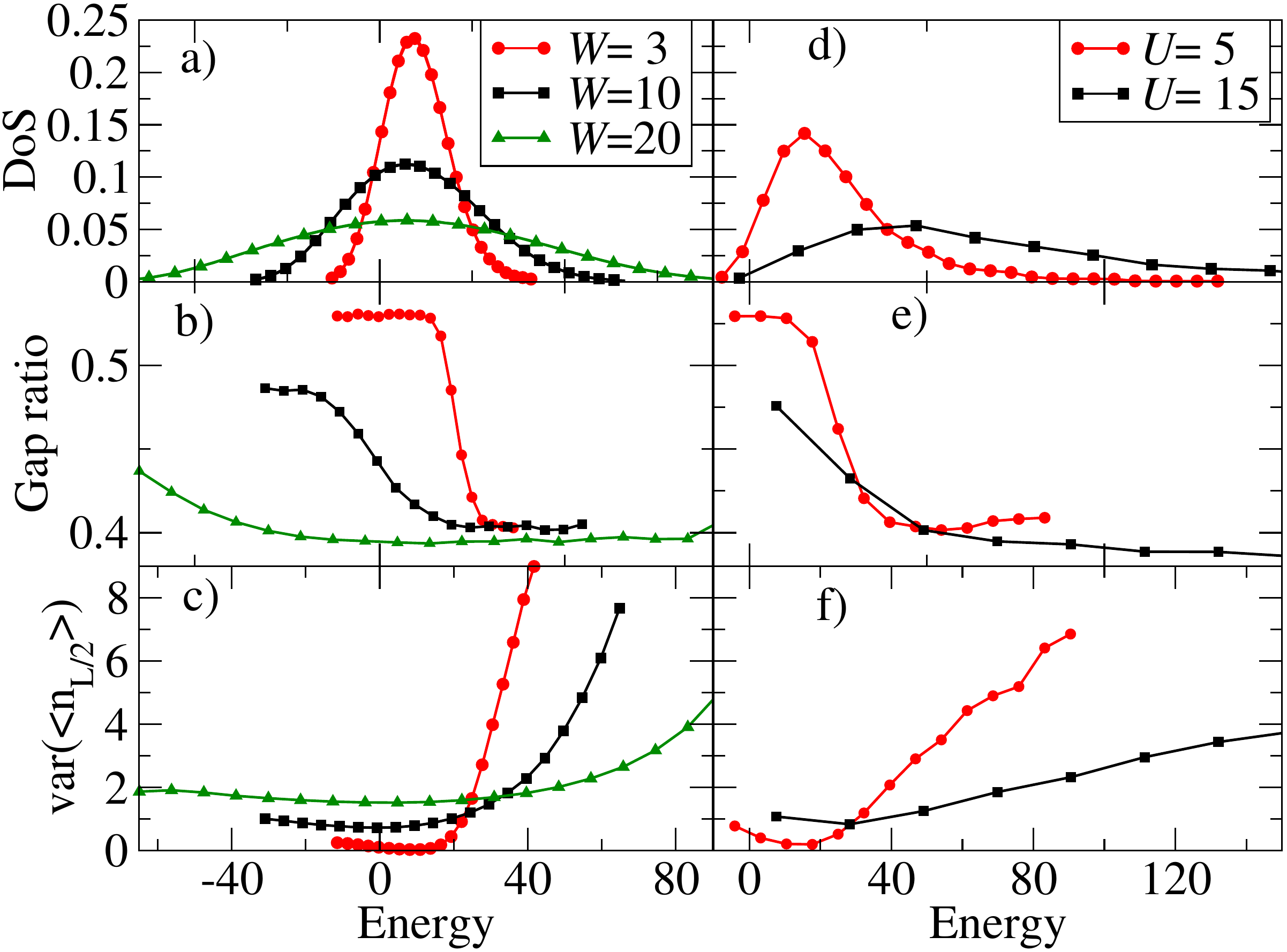}
\end{center}
\caption{ \label{fig: DOS}Density of states -- panels a) and d), mean ratio of consecutive spacings $\overline{r}$ as 
function of energy -- panels b) and e) and 
the variance of the occupation of the central $L/2$ site $\mathrm{var}(\langle n_{L/2}\rangle)$ as function of energy
-- panels c) and f) 
for the models with random 
chemical potential and $U=1$ (left column) and for  random interactions (right column). The disorder 
amplitudes are given in the figure. 
A system of  $N=9$ bosons in $L=6$ lattice sites (500 realizations averaged) is analyzed.
}
\end{figure}

\section{Interactions as a perturbation}
\label{pertu}

The perturbative work \cite{Basko06} established the existence of many--body localization
 for  a system of interacting fermions and the results were extended to  MBL of bosons  \cite{Aleiner09}.
The key element is the ratio, ${\cal R}$,  of the matrix element between 
two states localized at different sites
to the difference in their energies.  The ratio  larger than unity leads to 
delocalization, whereas the states remain localized for a small ${\cal R}$. This argument, used originally for
discussing Anderson transition, may be applied also to MBL. It also
lies  at the heart of the renormalization group treatment 
of MBL transition \cite{Vosk2015,Potter2015}.

It is straightforward to adapt this reasoning to the system with  random on--site potential 
(\ref{eq: BH_random_potential}). 
At $U=0$, it reduces to Anderson model --  working in 
 its single particle localized basis, we assume 
the localization/delocalization transition happens when
the energy mismatch between energies of initial states $\epsilon_i, \, \epsilon_j$ and final states  
$\epsilon_k, \, \epsilon_l$
becomes comparable with the coupling  $U_{ij,kl}$  between those states which stems from the on--site
interaction term in 
the Hamiltonian (\ref{eq: BH_random_potential}), so that
 \begin{equation}
  \displaystyle{
 { \cal R }  =  U_{ij,kl} /|\epsilon_i +\epsilon_j - \epsilon_l -\epsilon_k | \approx 1.
  } 
  \label{eq: loc-deloc}
  \end{equation}
The annihilation operator, $a_i$, associated with the Wannier basis state on site $i$ can
be written as a combination of operators
annihilating particles in single--particle localized states, $b_j$
\begin{equation}
  \displaystyle{
 a_i = \sum_j \varphi^j_i b_j,
  } 
  \label{eq: loc-deloc22}
  \end{equation}
where the coefficients $\varphi^j_i$ decay exponentially with the distance between sites $i$ and $j$.
Now, the single--particle part of (\ref{eq: BH_random_interactions}) can be written as 
\begin{equation}
  \displaystyle{
 H_0 = \sum_j \epsilon_j b^{\dag}_j b_j
  } 
  \label{eq: loc-deloc1}
  \end{equation}
and the interaction part becomes
  \begin{equation}
  \displaystyle{
 H_1 = \frac{1}{2}U\sum_{i,j_1,j_2,j_3,j_4} \varphi^{j_1}_i \varphi^{j_2}_i \varphi^{j_3}_i \varphi^{j_4}_i b^{\dag}_{j_1} 
 b^{\dag}_{j_2} b_{j_3} b_{j_4}.
  } 
  \label{eq: loc-deloc2}
  \end{equation}
  The exponential decay of $\varphi^j_i$ 
 with the distance $|i-j|$  implies that the terms 
 in (\ref{eq: loc-deloc2}) 
 can be organized order by order considering the index sum $S= |j-j_1|+|j-j_2|+|j-j_3|+|j-j_4|$.
 The zero order term $S=0$ reads 
 $\frac{1}{2}U  \sum_{i} (\varphi^i_i)^4 n^{b}_i(n^{b}_i -1)$ and corresponds to  a mere shift of energies  
 of the eigenstates of the system due to the interactions. The first order terms
 $S=1$ are of 
 the form of the density--induced tunnelings $b^{\dag}_{i} n^b_i b_{i+1}$ and  are smaller than the 
 $S=0$ terms by a 
 factor $\varphi^{i+1}_i/ \varphi^i_i$. 
 This classification may be continued to higher order terms.
  
  The question is whether the condition  (\ref{eq: loc-deloc}) together with the form of the interaction--induced 
 terms can be used to get some qualitative understanding of the observed ergodic to MBL transition in the interacting system.
  Consider the system with random a on--site potential and its density of states (DoS) -- Fig.~\ref{fig: DOS}a. 
  At high energies, 
 the density of states is small, therefore, the energy mismatches between the states that can be coupled by 
 the interaction
 (\ref{eq: loc-deloc2}) are sufficiently large for the system to remain localized even in the presence of interactions. 
 Consequently
 $\overline{r} \approx 0.4$ in that region of the spectrum -- Fig.~\ref{fig: DOS}b. 
 Now, as the energy decreases, the DoS grows larger and the energy differences 
 between states coupled by the off-diagonal part of (\ref{eq: loc-deloc2}) become comparable
 with the coupling. For $W=3$, states in this part of the spectrum are ergodic --
 $\overline{r} \approx 0.53$. On the other hand, for
 $W=10$, the coupling is not strong enough,
 $\overline{r} \approx 0.45$ and the system is in the intermediate regime.
 
 However, at the smallest energies, the DoS is low again -- so why the does system
 remain ergodic (or closer to the ergodic phase)
 even though the 
 energy mismatches seem to be bigger again? First, let us note, that the DoS is slightly asymmetric 
 due to  the quadratic nature of the interaction 
 term. Consider a state at the bottom of the spectrum -- it necessarily has large occupations of 
 single--particle orbitals localized around sites
 with a large negative chemical potential. The  $S=0$ term $n^{b}_i(n^{b}_i -1)$ shifts
 energy of this state upwards, to a region of spectrum where the DoS
 is higher. Now consider a state at the top of the spectrum, its energy is again 
 increased by the interactions but now the state is 
 shifted towards the high end of the spectrum, where the  DoS is low. Therefore, the $S=0$ term is  
 the first source of an asymmetry between lower and higher 
 parts of the spectrum. However, the differences between eigenstates in those regions are much 
 more pronounced which is well captured by $\mathrm{var}(\langle n_{L/2}\rangle)$
-- variance (with respect to different disorder realizations) of the average occupation of 
the central site of the lattice, which is
shown in Fig.~\ref{fig: DOS}c.
 The $\mathrm{var}(\langle n_{L/2}\rangle)$ increases rapidly at large energies and renders 
 the system localized. {Conversely, at low energies, fluctuations of the number of the 
 particles are smaller and the interactions delocalize the eigenstates easily.}
 
 The system with random interactions -- see the Fig.~\ref{fig: DOS}d-f is not
  straightforwardly treatable with the same perturbative
 analysis as the single--particle physics at $U=0$ is delocalized. However, the general results of
 the reasoning for the random chemical potential 
 are the same - the DoS is much more asymmetric and states at high energies are localized.
 Moreover, $\mathrm{var}(\langle n_{L/2}\rangle)$ looks quantitatively the 
 same and shows that also for the random interactions states at the lower parts of spectrum
 have a better chance of being ergodic.
 
 Concluding, the inverted localization character of states (with low energy states being 
 extended and high energy states tending to be localized) in the  Bose--Hubbard model 
 stems from the possible higher than one occupation of lattice 
 sites ($\mathrm{var}(\langle n_{L/2}\rangle) > 1$) and the fact that the interaction energy
 increases quadratically with the site occupation.
  
\section{Level statistics}
\label{statu}

The spectral statistics are a useful probe of the MBL transition which we have already seen
in Sec.~\ref{seclocal} discussing the gap ratio. Let us now
concentrate on a more traditional level statistics, the distribution of spacings. 
A generic ergodic system  is characterized by Wigner-Dyson statistics characteristic 
for GOE matrices, whereas for an integrable system (i.e. also in the MBL phase where a 
complete set of LIOMs exists \cite{Serbyn13b,Huse14,Imbrie17}) the
 Poissonian statistics is appropriate \cite{Haake}. 
The intermediate statistics in the context of  ergodic to many--body localized transition 
is addressed in \cite{Serbyn16} on an example of a
Heisenberg $XXZ$ spin chain. Serbyn and Moore postulated that the transition might be 
described by the so called plasma model \cite{Serbyn16}. 
More precisely, using a random walk approach in a space of
Hamiltonians generated by different realizations of disorder \cite{Chalker96}, 
with mean field based assumptions on the correlation functions for these Hamiltonians, as well 
as assuming fractal scaling  to the matrix elements of 
local operators Serbyn and Moore obtain a power law scaling for the disorder inducing term of the Hamiltonian
(they explicitly consider XXZ spin model). This allows them, in turn, to map their  model 
onto the plasma  model for level statistics \cite{Kravtsov95}. The model predicts
 the level spacing distribution $P(s)$ for large $s$ 
that interpolates between the exponential and 
the Gaussian tail. Assuming also a plausible power law repulsion at small spacings, one arrives at  \cite{Serbyn16}
\begin{equation}
P(s) \propto s^{\beta}\mathrm{e}^{-C_{\gamma,\beta} s^{2-\gamma}} \label{plasmamodel}
\end{equation}
with $\beta$ and $\gamma$ being the parameters of the plasma model. It is shown to well  describe the
level spacing distribution for the studied spin system across the transition. 

 Level spacing distributions were also used by us \cite{Sierant17} for randomly interacting bosons
 to estimate the position of the critical 
region between the ergodic and localized phases. We found that the distribution of spacings
was well described by the plasma model distribution between GOE 
and Poisson limits. In fact, we have identified two regimes; the generalized semi-Poisson \cite{Bogomolny99}
regime close to the MBL phase corresponding to $\gamma=1$ and variable $\beta$
followed by the transition to GOE for $\beta=1$ and $\gamma$ decreasing to zero.
In the following, we discuss mainly the spacing distribution for the random chemical potential
model (\ref{eq: BH_random_potential}) and make some comments on the random interactions model.

\begin{figure}[ht]
\begin{center}\includegraphics[width=0.8\columnwidth]{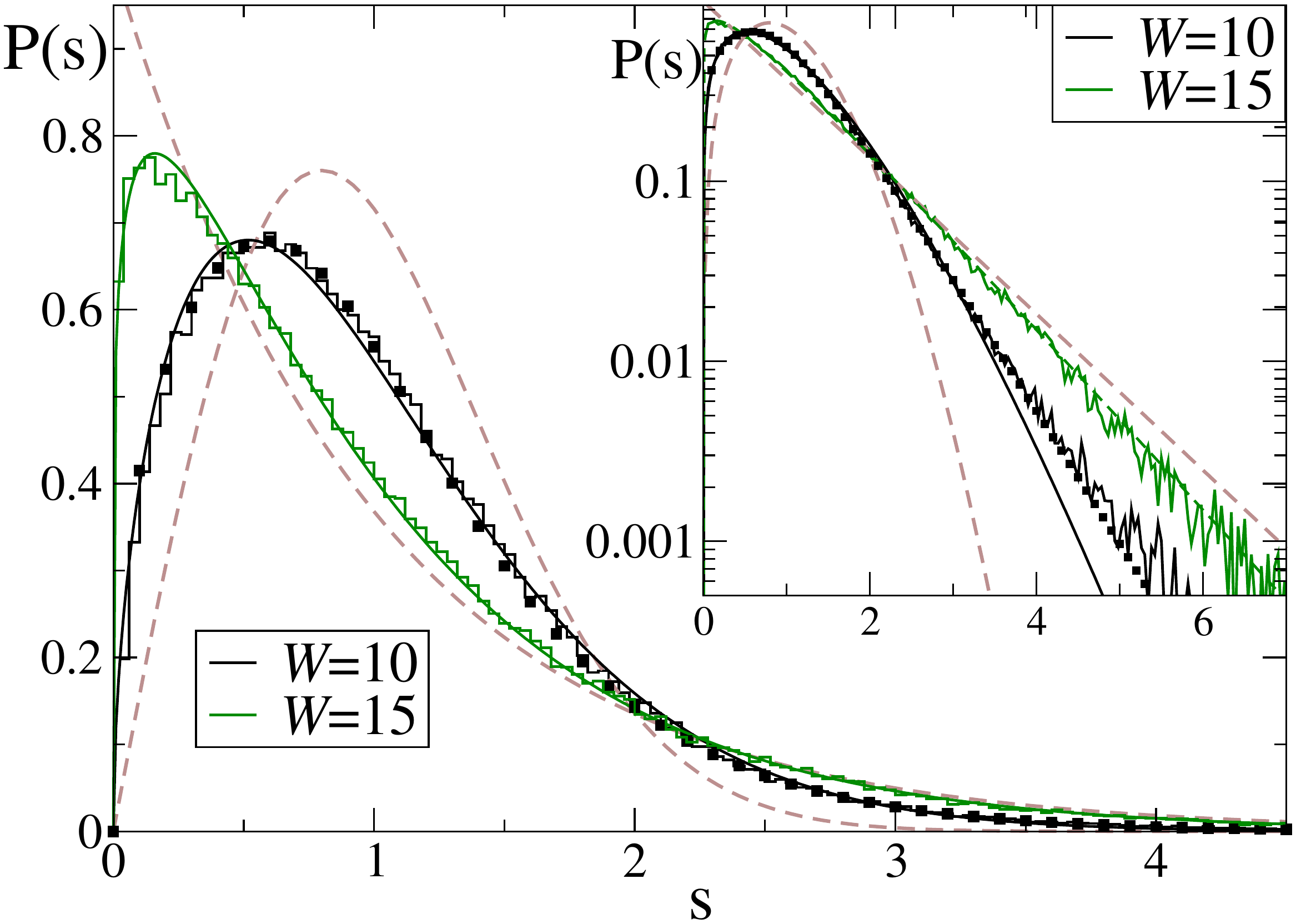}
\end{center}
\caption{Level spacings distributions for the system with random chemical potential for 
$N=12$ bosons on $L=8$ sites with $U=1$. Data for $W=1$ and $W=30$ are well reproduced by the Wigner's 
surmise formula and Poisonian statistics (brown dashed lines) and are not displayed to ensure better 
visibility of data for $W=10,15$.
Those are fitted with the plasma model distribution and semi--Poissonian statistics 
respectively (solid lines). Finally, the squares
correspond to our effective formula (\ref{eq: erf}). The inset presents the same data in the
lin-log scale showing that the plasma model fails to reproduce
the exponential tail of the numerical data despite a seemingly good fit of the bulk (main panel).} 
\label{fig: spacingsW}
\end{figure}

After performing the necessary unfolding of energy levels \cite{Haake} we obtain the distribution
of spacings between neighboring energy levels $P(s)$ with the mean level spacing equal to unity -
the results for a  a random chemical potential  are presented in Fig.~\ref{fig: spacingsW}. 
The level spacing $P(s)$  at small disorder ($W=1 $) is well approximated by the Wigner's
surmise \cite{bgs84}
$P(s) = \frac{\pi}{2} s \mathrm{e}^{-\frac{\pi}{4}s^2}$ which confirms the ergodic behavior of the system. 
At large values of disorder ($W=25$), deep in the MBL phase, the system is fully integrable 
\cite{Serbyn13b, Huse14}, and the resulting spacing distribution is Poissonian $P(s) = \mathrm{e}^{-s}$.

It is notable that the transition between the two limiting distributions follows the similar 
pattern to that observed before \cite{Serbyn16,Sierant17}.
In the transition region, we observe a two stage process as in \cite{Sierant17}. However, at smaller values of
disorder we find out, quite surprisingly,
that the proposed plasma model \cite{Serbyn16}, while nicely reproducing the bulk of
the distribution for $W=10$ -- see Fig.~\ref{fig: spacingsW}, does not reproduce our data well in the tail of 
the distribution, as 
shown in the inset of Fig.~\ref{fig: spacingsW}. For fitted values of $\beta$ and $\gamma$ the plasma model
distribution decays faster than exponentially $\gamma=0.59<1$
while the numerically obtained data reveal an exponential tail. Forcing $\gamma=1$ to get the agreement in 
the tail leads to a poor comparison of numerics and plasma model distribution for small and moderate spacings. 
To resolve this issue we fit the data for $W=10$ with the formula
\begin{equation}
 P(s) = s^{\beta} (C_1 +C_2 \mathrm{Erf}(C_3(s-s_0))\mathrm{e}^{-\alpha s},
 \label{eq: erf}
\end{equation}
in such a way, that $\beta$ and $\alpha$ are fixed by the limiting behavior of $P(s)$ at small and large $s$, 
respectively, two of the $C_{1,2,3}$ constants are fixed
by the requirement that $\langle 1 \rangle=1=\langle s\rangle$ and the remaining one and $s_0$ are fitted. The fit of (\ref{eq: erf}) 
reproduces both the tail and the bulk of the level
spacing distribution more accurately than the plasma model prediction. {Let us stress, that the formula
(\ref{eq: erf}) is not a result of a deeper theory, but rather a heuristic formula which grasps effectively all 
essential features of the level spacing distribution.}

The deviation from the plasma model occurs close to the delocalized regime. Importantly, after a close inspection, we have 
observed exactly the same behavior for the model with random interactions.
While the distribution of the bulk (small or intermediate $s$) of the spacing distribution was well captured 
by the plasma model as reported by us \cite{Sierant17}, the large spacing tail remained exponential and the fits
with the proposed distribution (\ref{eq: erf}) were clearly superior (since the data resemble that of 
Fig.~\ref{fig: spacingsW} we do not reproduce them).

At larger disorder strengths, level spacing distributions of both considered systems
(\ref{eq: BH_random_potential}), (\ref{eq: BH_random_interactions}) are well described by 
the generalized semi--Poison distribution 
$P(s) \propto s^{\beta}\mathrm{e}^{-(\beta+1)s}$.

Concluding, the level statistics for the Bose--Hubbard model with random on--site chemical
potential or with random interactions
reveal the ergodic to MBL transition and the level spacings distributions in the intermediate 
regime are similar to XXZ--Heisenberg 
spin chain \cite{Serbyn16}. However, the 2-parameter plasma model is insufficient to capture the
level spacings in the critical 
regime close to the ergodic phase (as for $W=10$) because of the exponentially decaying tails
of numerically obtained $P(s)$.

\section{Fate of metastable states in presence of disorder}
\label{fateu}

The previous sections demonstrate that the disordered Bose--Hubbard models possess the characteristic features 
of MBL systems. 
In particular, following the time evolution of the initial density--wave states one observes ergodicity breaking.
 However, it has been shown that \cite{Carleo12}
strong repulsive interactions of bosons lead to dynamical constraints which slow down
thermalization of the system
as soon as it is prepared in a highly excited inhomogeneous initial state.
In this section, we compare this mechanism to the disorder induced ergodicity breaking.  

The overall thermalization rate of the clean system depends on the population of 
high--energy excitations, which, at strong 
interactions, is associated with states having sites occupied by more than a single boson.
\begin{figure}[h]
\begin{center}\includegraphics[width=0.8\columnwidth]{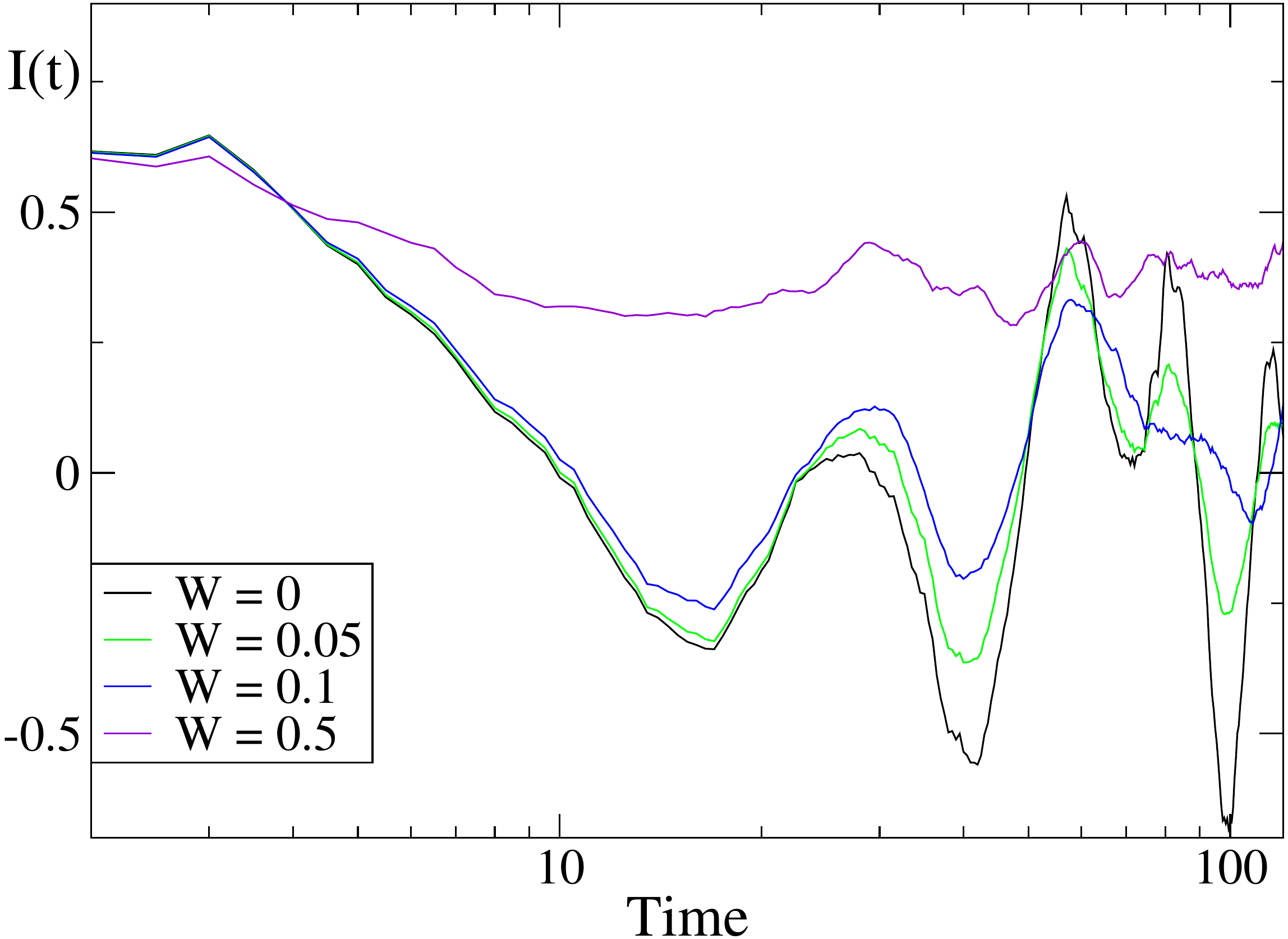}
\end{center}
\caption{ \label{fig: meta2} (color online) $I(t)$ for various disorder strengths
(averaged over $30$ disorder realizations) for $N=8$ bosons on $L=8$ 
sites at $U = 10$. 
With growing disorder strength, the oscillations amplitude 
gets smaller and a non--zero average value of imbalance is obtained -- 
already relatively small disorder $W=0.5$ leads to significant
ergodicity breaking and $I_{stat}\approx0.36$.
}
\end{figure}
\begin{figure}[ht]
\begin{center}
\includegraphics[width=0.8\columnwidth]{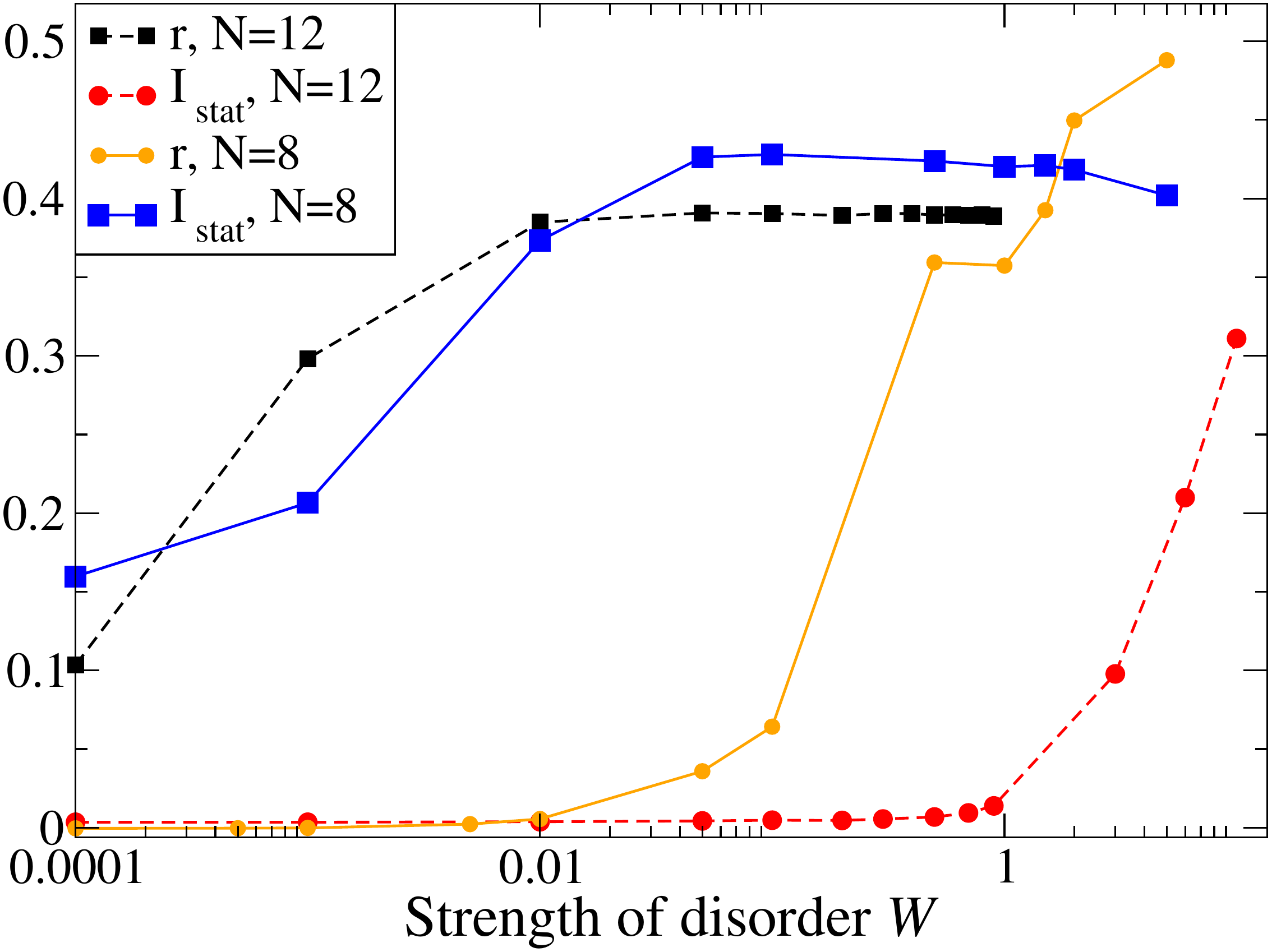}
\end{center}
\caption{ \label{fig: meta1} Change in $\bar{r}$ and stationary value of imbalance resulting 
from introducing disorder to the system. 
At small $W$ the $\overline{r}$ value is smaller than value characteristic for uncorrelated 
Poissonian energy levels -- once $\bar{r}\approx 0.4$
is attained, non--zero stationary value of imbalance is observed and the system is no longer ergodic.
The parameters are $U=20$ for $N=12$ and $U=10$ for $N=8$.
}
\end{figure}
For instance, doublons from the density--wave state $|DW_{20}\rangle$ (at filling $\nu = 1$) 
very slowly decay at large $U$ 
being incapable of moving and restoring translational symmetry. That was quantified in
\cite{Carleo12} by the relaxation time 
$\tau_R$ - the smallest time at which the local density acquires its equilibrium value. $\tau_R$ 
was found to be increasing with the interaction strength and 
with the growing system size.

In order to demonstrate that the dynamical trapping described above and MBL induced by disorder are two physically
distinct phenomena, we consider a 
system of $N=8$ bosons on $L=8$ sites with strong interactions $U=10$ and gradually increase disorder strength $W$. 
The relaxation time of the system is $\tau_R \approx 10$. After this time $I(t)$ oscillates around zero and the system
thermalizes.
The imbalance $I(t)$ calculated for different disorder strengths is presented in Fig.~\ref{fig: meta2}.
Clearly, even though the time evolution at small times 
is similar for different , and hence $\tau_R$ does not change drastically in the interval $W\in [0, 0.3 ]$,
the long time evolution is affected -- 
the oscillations of $I(t)$ (disorder averaged) are smaller and already for $W=0.1$ a non--zero stationary 
value of imbalance builds up.
The dependence of stationary values of imbalance on the disorder strength together with the corresponding $\overline{r}$ values 
are presented in Fig. \ref{fig: meta1}.
Clearly, at larger $W$ a non--zero stationary value of imbalance is observed, moreover it happens only at disorder strengths 
that correspond to $\bar{r}\approx 0.4$ -- i.e in the MBL regime. 

Therefore, for very small disorder strengths the slow down of relaxation due to metastable states at large
$U$ prevails. However, an addition
of even a small disorder to this strongly interacting system affects its dynamics severely and leads to a much more robust 
mechanism of ergodicity breaking.

\section{Conclusions}
\label{concu}

We have presented a comprehensive analysis of many-body localization for a system of interacting bosons
in  a lattice in the presence of disorder. We considered both the random chemical potential as well as 
we revisited the case of random interactions. Both systems can be realized experimentally in optical lattices.

The treatment of interacting bosons in a lattice is technically
more difficult than spin=1/2 or fermionic systems due to the possibility of multiple occupations of
individual sites. This, in principle could lead to profound differences w.r.t. fermions/spins as the 
local Hilbert space might be viewed as an additional synthetic dimension.
However, as found above, the 
 repulsive interactions between bosons limit multiple occupancies to uninteresting 
high energy physics.  
For the majority of states, at the intermediate energies the character of
MBL observed  is similar to that of 
spins and fermions. In particular, MBL may be evidenced by a long-time behavior of the imbalance of appropriately
prepared inhomogeneous initial states. In this respect bosonic physics is 
richer, allowing for a preparation of different states, possibly differing in energy.
This may allow one to experimentally study the energy dependence of the transition between ergodic and MBL phases 
when the disorder is increased. 
This is particularly interesting as we have shown that the system reveals an
apparent localization (mobility) edge. Moreover this ``edge'' is inverted in a sense that localized states lay 
higher in energy than the extended ergodic states that occupy lower energy sector. 
Let us stress that it is the bosonic nature of the models 
which allows for initial density wave states at significantly different energies. An experiment
in which the time evolution of the imbalance starting from the $|DW_{21} \rangle$ and 
$|DW_{30}\rangle$ density--wave states would verify the existence of the apparent mobility edges.

In the critical quantum 
regime between ergodic and localized phases (in our necessarily finite size systems studies) we 
observe an algebraic decay of the imbalance in agreement with the subdiffusive character predicted in a general 
one-dimensional
renormalization group theory  \cite{Vosk2015,Potter2015}. 

In the case of the 
revisited random
interactions, the thorough investigation of the critical region 
characterized by subdiffusion for small system sizes leads us to  a better estimation of the 
stationary imbalance.
The results about the mobility edge are confirmed for larger system size. The direct comparison 
with the random on--site potential model allows us to speculate that the underlying mechanism 
of MBL is similar in both systems.

The detailed analysis of level spacing distributions in the transition regime revealed that 
the so called plasma model \cite{Serbyn16} fails to reproduce the behavior of the tail of the distribution
despite faithfully reproducing the bulk. On the other hand the similarity 
of spin and boson level statistics in the transition regime suggests a significant level of universality in 
the transition between ergodic - many-body localized phases and calls for a separate analysis. Such an analysis is in progress.

\section{Acknowledgement} 
We are grateful to Dominique Delande for illuminating discussions throughout the course of this work. 
We acknowledge also useful conversations with Fabian Alet and Antonello Scardicchio. 
{We thank Jesse Mumford for reading and correcting the grammar in the manuscript.}
This research was performed within 
project   No.2015/19/B/ST2/01028  financed by  National Science Centre (Poland). Support by PL-Grid Infrastructure and by EU via project
QUIC (H2020-FETPROACT-2014 No. 641122) is  also acknowledged.

\section{References}

\begin{thebibliography}{10}
\expandafter\ifx\csname url\endcsname\relax
  \def\url#1{{\tt #1}}\fi
\expandafter\ifx\csname urlprefix\endcsname\relax\def\urlprefix{URL }\fi
\providecommand{\eprint}[2][]{\url{#2}}

\bibitem{Anderson58}
Anderson P~W 1958 {\em Phys. Rev.\/} {\bf 109} 1492

\bibitem{Altshuler79}
Altshuler B~L and Aronov A~G 1979 {\em Solid State Commun.\/} {\bf 30} 115

\bibitem{Altshuler80}
Altshuler B~L, Aronov A~G and Lee P~A 1980 {\em Phys. Rev. Lett.\/} {\bf 44}
  1288

\bibitem{Fukuyama80}
Fukuyama H 1980 {\em J. Phys. Soc. Jpn.\/} {\bf 48} 2169

\bibitem{Fleishman1980}
Fleishman L and Anderson P~W 1980 {\em Phys. Rev. B\/} {\bf 21}(6) 2366--2377
  \urlprefix\url{https://link.aps.org/doi/10.1103/PhysRevB.21.2366}

\bibitem{Shepelyansky94}
Shepelyansky D~L 1994 {\em Phys. Rev. Lett.\/} {\bf 73} 2607

\bibitem{Damski2003}
Damski B, Zakrzewski J, Santos L, Zoller P and Lewenstein M 2003 {\em Phys.
  Rev. Lett.\/} {\bf 91} 080403

\bibitem{Basko06}
Basko D, Aleiner I and Altschuler B 2006 {\em Ann. Phys. (NY)\/} {\bf 321} 1126

\bibitem{Huse14}
Huse D~A, Nandkishore R and Oganesyan V 2014 {\em Phys. Rev. B\/} {\bf 90}(17)
  174202 \urlprefix\url{http://link.aps.org/doi/10.1103/PhysRevB.90.174202}

\bibitem{Rahul15}
Nandkishore R and Huse D~A 2015 {\em Ann. Rev. Cond. Mat. Phys.\/} {\bf 6} 15

\bibitem{Abanin17}
Abanin D~A and {Papi\'c} Z 2017 {\em Annalen der Physik\/} {\bf 529} 1700169
  1700169 \urlprefix\url{http://dx.doi.org/10.1002/andp.201700169}

\bibitem{Alet17}
{Alet} F and {Laflorencie} N 2017 {\em ArXiv e-prints:\/}  1711.03145

\bibitem{Schreiber15}
Schreiber M, Hodgman S~S, Bordia P, L\"uschen H~P, Fischer M~H, Vosk R, Altman
  E, Schneider U and Bloch I 2015 {\em Science\/} {\bf 349} 7432

\bibitem{Kondov15}
Kondov S~S, McGehee W~R, Xu W and DeMarco B 2015 {\em Phys. Rev. Lett.\/} {\bf
  114}(8) 083002
  \urlprefix\url{https://link.aps.org/doi/10.1103/PhysRevLett.114.083002}

\bibitem{White09}
White M, Pasienski M, McKay D, Zhou S~Q, Ceperley D and DeMarco B 2009 {\em
  Phys. Rev. Lett.\/} {\bf 102}(5) 055301
  \urlprefix\url{https://link.aps.org/doi/10.1103/PhysRevLett.102.055301}

\bibitem{Choi16}
Choi J~y, Hild S, Zeiher J, Schau{\ss} P, Rubio-Abadal A, Yefsah T, Khemani V,
  Huse D~A, Bloch I and Gross C 2016 {\em Science\/} {\bf 352} 1547--1552

\bibitem{Tang15}
Tang B, Iyer D and Rigol M 2015 {\em Phys. Rev. B\/} {\bf 91}(16) 161109
  \urlprefix\url{http://link.aps.org/doi/10.1103/PhysRevB.91.161109}

\bibitem{Sierant17}
Sierant P, Delande D and Zakrzewski J 2017 {\em Phys. Rev. A\/} {\bf 95}(2)
  021601 \urlprefix\url{https://link.aps.org/doi/10.1103/PhysRevA.95.021601}

\bibitem{Sierant17b}
Sierant P, Delande D and Zakrzewski J 2017 {\em Acta Phy. Polon. B\/} {\bf in
  press} xxxx \urlprefix\url{https://arxiv.org/abs/1707.08845}

\bibitem{Zhou10}
Zhou S~Q and Ceperley D~M 2010 {\em Phys. Rev. A\/} {\bf 81}(1) 013402
  \urlprefix\url{https://link.aps.org/doi/10.1103/PhysRevA.81.013402}

\bibitem{Bakr2009}
Bakr W~S, Gillen J~I, Peng A, F\"{o}lling S and Greiner M 2009 {\em Nature\/}
  {\bf 462} 74--77 \urlprefix\url{http://dx.doi.org/10.1038/nature08482}

\bibitem{Gimperlein05}
Gimperlein H, Wessel S, Schmiedmayer J and Santos L 2005 {\em Phys. Rev.
  Lett.\/} {\bf 95} 170401

\bibitem{Lueschen17}
L\"uschen H~P, Bordia P, Scherg S, Alet F, Altman E, Schneider U and Bloch I
  2017 {\em Phys. Rev. Lett.\/} {\bf 119}(26) 260401
  \urlprefix\url{https://link.aps.org/doi/10.1103/PhysRevLett.119.260401}

\bibitem{Ancilotto18}
{Ancilotto} F, {Rossini} D and {Pilati} S 2018 {\em ArXiv e-prints\/}
  (\textit{Preprint} \eprint{1801.05596})

\bibitem{Zakrzewski18}
{Zakrzewski} J and {Delande} D 2018 {\em ArXiv e-prints\/} (\textit{Preprint}
  \eprint{1802.02525})

\bibitem{Carleo12}
Carleo G, Becca F, Schiro M and Fabrizion M 2012 {\em Scientific Reports\/}
  {\bf 2}(243) \urlprefix\url{http://dx.doi.org/10.1038/srep00243}

\bibitem{Deutsch91}
Deutsch J~M 1991 {\em Phys. Rev. A\/} {\bf 43}(4) 2046--2049
  \urlprefix\url{https://link.aps.org/doi/10.1103/PhysRevA.43.2046}

\bibitem{Srednicki94}
Srednicki M 1994 {\em Phys. Rev. E\/} {\bf 50}(2) 888--901
  \urlprefix\url{https://link.aps.org/doi/10.1103/PhysRevE.50.888}

\bibitem{Cohen16}
Cohen D, Yukalov V~I and Ziegler K 2016 {\em Phys. Rev. A\/} {\bf 93}(4) 042101
  \urlprefix\url{https://link.aps.org/doi/10.1103/PhysRevA.93.042101}

\bibitem{Luitz15}
Luitz D~J, Laflorencie N and Alet F 2015 {\em Phys. Rev. B\/} {\bf 91}(8)
  081103 \urlprefix\url{https://link.aps.org/doi/10.1103/PhysRevB.91.081103}

\bibitem{Luitz16}
Luitz D~J, Laflorencie N and Alet F 2016 {\em Phys. Rev. B\/} {\bf 93}(6)
  060201 \urlprefix\url{http://link.aps.org/doi/10.1103/PhysRevB.93.060201}

\bibitem{Park86}
Jun~Park T and Light J 1986 {\em J. Chem. Phys.\/} {\bf 85} 5870--5876

\bibitem{Agarwal15}
Agarwal K, Gopalakrishnan S, Knap M, M\"uller M and Demler E 2015 {\em Phys.
  Rev. Lett.\/} {\bf 114}(16) 160401
  \urlprefix\url{https://link.aps.org/doi/10.1103/PhysRevLett.114.160401}

\bibitem{Lev15}
Bar~Lev Y, Cohen G and Reichman D~R 2015 {\em Phys. Rev. Lett.\/} {\bf 114}(10)
  100601
  \urlprefix\url{https://link.aps.org/doi/10.1103/PhysRevLett.114.100601}

\bibitem{Torres-Herrera15}
Torres-Herrera E~J and Santos L~F 2015 {\em Phys. Rev. B\/} {\bf 92}(1) 014208
  \urlprefix\url{https://link.aps.org/doi/10.1103/PhysRevB.92.014208}

\bibitem{Vosk2015}
Vosk R, Huse D~A and Altman E 2015 {\em Phys. Rev. X\/} {\bf 5}(3) 031032
  \urlprefix\url{https://link.aps.org/doi/10.1103/PhysRevX.5.031032}

\bibitem{Potter2015}
Potter A~C, Vasseur R and Parameswaran S~A 2015 {\em Phys. Rev. X\/} {\bf 5}(3)
  031033 \urlprefix\url{https://link.aps.org/doi/10.1103/PhysRevX.5.031033}

\bibitem{Khemani17}
Khemani V, Lim S~P, Sheng D~N and Huse D~A 2017 {\em Phys. Rev. X\/} {\bf 7}(2)
  021013 \urlprefix\url{https://link.aps.org/doi/10.1103/PhysRevX.7.021013}

\bibitem{Griffiths69}
Griffiths R~B 1969 {\em Phys. Rev. Lett.\/} {\bf 23}(1) 17--19
  \urlprefix\url{https://link.aps.org/doi/10.1103/PhysRevLett.23.17}

\bibitem{Vojta10}
Vojta T 2010 {\em J. Low Temp. Phys.\/} {\bf 161} 299--323

\bibitem{Agarwal16}
Agarwal K, Altman E, Demler E, Gopalakrishnan S, A~Huse D and Knap M 2017 {\em
  Annalen der Physik\/} {\bf 529} 201600326

\bibitem{Vidal03}
Vidal G 2003 {\em Phys. Rev. Lett.\/} {\bf 91}(14) 147902
  \urlprefix\url{http://link.aps.org/doi/10.1103/PhysRevLett.91.147902}

\bibitem{Vidal04}
Vidal G 2004 {\em Phys. Rev. Lett.\/} {\bf 93}(4) 040502
  \urlprefix\url{http://link.aps.org/doi/10.1103/PhysRevLett.93.040502}

\bibitem{Zakrzewski09}
Zakrzewski J and Delande D 2009 {\em Phys. Rev. A\/} {\bf 80}(1) 013602
  \urlprefix\url{http://link.aps.org/doi/10.1103/PhysRevA.80.013602}

\bibitem{Schollwoeck11}
Schollwoeck 2011 {\em Ann. Phys. (NY)\/} {\bf 326} 96

\bibitem{Serbyn13a}
Serbyn M, Papi\ifmmode~\acute{c}\else \'{c}\fi{} Z and Abanin D~A 2013 {\em
  Phys. Rev. Lett.\/} {\bf 110}(26) 260601
  \urlprefix\url{http://link.aps.org/doi/10.1103/PhysRevLett.110.260601}

\bibitem{Mott}
Mott N~F 1990 {\em Metal-Insulator Transitions, 2nd Edition\/} (Taylor and
  Francis Ltd., London)

\bibitem{Naldesi16}
Naldesi P, Ercolessi E and Roscilde T 2016 {\em SciPost Phys.\/} {\bf 1}(1) 010
  \urlprefix\url{https://scipost.org/10.21468/SciPostPhys.1.1.010}

\bibitem{Lin17}
{Lin} S~H, {Sbierski} B, {Dorfner} F, {Karrasch} C and {Heidrich-Meisner} F
  2017 {\em ArXiv e-prints\/}  1707.06759

\bibitem{Oganesyan07}
Oganesyan V and Huse D~A 2007 {\em Phys. Rev. B\/} {\bf 75}(15) 155111
  \urlprefix\url{http://link.aps.org/doi/10.1103/PhysRevB.75.155111}

\bibitem{Atas13}
Atas Y~Y, Bogomolny E, Giraud O and Roux G 2013 {\em Phys. Rev. Lett.\/} {\bf
  110}(8) 084101
  \urlprefix\url{http://link.aps.org/doi/10.1103/PhysRevLett.110.084101}

\bibitem{Serbyn13b}
Serbyn M, Papi\ifmmode~\acute{c}\else \'{c}\fi{} Z and Abanin D~A 2013 {\em
  Phys. Rev. Lett.\/} {\bf 111}(12) 127201
  \urlprefix\url{http://link.aps.org/doi/10.1103/PhysRevLett.111.127201}

\bibitem{Imbrie17}
Imbrie J~Z, Ros V and Scardicchio A 2017 {\em Annalen der Physik\/} {\bf 529}
  201600278

\bibitem{Pal10}
Pal A and Huse D~A 2010 {\em Phys. Rev. B\/} {\bf 82}(17) 174411
  \urlprefix\url{http://link.aps.org/doi/10.1103/PhysRevB.82.174411}

\bibitem{Singh17}
Singh R and Shimshoni E 2017 {\em Annalen der Physik\/} {\bf 529} 1600309
  \urlprefix\url{http://dx.doi.org/10.1002/andp.201600309}

\bibitem{Haake}
Haake F 2010 {\em Quantum Signatures of Chaos\/} (Springer, Berlin)

\bibitem{Prozen94}
Prosen T and Robnik M 1994 {\em Journal of Physics A: Mathematical and
  General\/} {\bf 27} L459
  \urlprefix\url{http://stacks.iop.org/0305-4470/27/i=13/a=001}

\bibitem{Aleiner09}
Aleiner I~L, Altshuler B~L and Shlyapnikov G~V 2010 {\em Nat Phys\/} {\bf 6}
  900--904 \urlprefix\url{http://dx.doi.org/10.1038/nphys1758}

\bibitem{Serbyn16}
Serbyn M and Moore J~E 2016 {\em Phys. Rev. B\/} {\bf 93}(4) 041424
  \urlprefix\url{http://link.aps.org/doi/10.1103/PhysRevB.93.041424}

\bibitem{Chalker96}
Chalker J~T, Lerner I~V and Smith R~A 1996 {\em Phys. Rev. Lett.\/} {\bf 77}(3)
  554--557 \urlprefix\url{https://link.aps.org/doi/10.1103/PhysRevLett.77.554}

\bibitem{Kravtsov95}
Kravtsov V~E and Lerner I 1995 {\em J. Phys. A: Math. Gen.\/} {\bf 28} 3623

\bibitem{Bogomolny99}
Bogomolny E~B, Gerland U and Schmit C 1999 {\em Phys. Rev. E\/} {\bf 59}(2)
  R1315--R1318
  \urlprefix\url{https://link.aps.org/doi/10.1103/PhysRevE.59.R1315}

\bibitem{bgs84}
Bohigas O, Giannoni M~J and Schmit C 1984 {\em Phys. Rev. Lett.\/} {\bf 52}(1)
  1--4 \urlprefix\url{https://link.aps.org/doi/10.1103/PhysRevLett.52.1}

\end{thebibliography}
\providecommand{\newblock}{}

  \end{document}